\documentstyle[12pt]{article}

\begin{document}
{\sf \begin{center} \noindent {\Large \bf Cosmic kinematic dynamos in anisotropic non-singular universe}\\[3mm]

by \\[0.3cm]{\sl L.C. Garcia de Andrade}\\

\vspace{0.5cm} Departamento de F\'{\i}sica
Te\'orica -- IF -- Universidade do Estado do Rio de Janeiro-UERJ\\[-3mm]
Rua S\~ao Francisco Xavier, 524\\[-3mm]Cep 20550-003, Maracan\~a, Rio de Janeiro, RJ, Brasil\\[-3mm]
Electronic mail address: garcia@dft.if.uerj.br\\[-3mm]
\vspace{2cm} {\bf Abstract}
\end{center}
\paragraph*{}
A marginally excited cosmic kinematic dynamo is found in the
background of a non-singular anisotropic Kasner cosmological metric
solution of Einstein field equation of general relativity. The
magnetic field is not amplified but is frozen inside the universe.
Since a finite resistivity is assumed, a nonsingular flow velocity
is orthogonal to the magnetic field plane. The model presents an
inflationary phase. Magnetic field components are stretched along
the orthogonal planar directions while the flow is orthogonal to
this magnetic sheet. The self-induction magnetohydrodynamic field
equation in Kasner universe may be written in the form of a
polynomial in the time coordinate which yields a periodic flow.
\vspace{0.5cm} \noindent {\bf PACS numbers:}
\hfill\parbox[t]{13.5cm}{02.40.Hw-Riemannian geometries}

\newpage
 \section{Introduction}
 Cosmic and galactic dynamos have been widely investigated in recent
 years \cite{1,2,3}. More recently Brandenburg et al \cite{4} have consider a turbulent
 solution of the Einstein field equation representing the Friedmann-Robertson-Walker (FRW) early
 universe cosmic dynamo. They investigate the numerical solution of
 hydromagnetic turbulence of primordial fields. Though the authors considered that though the magnetic
 fields generate local bulk motion, and that it is in large scales consistent with homogeneity and isotropy
 of the cosmological model, here we argue that the anisotropic models could be consider exactly if one drops the exigence of
 large scales. In this brief report
 we present a very simple analytical solution of kinematic dynamo MHD equations ,
 which later on can be transformed on a more accurate physical
 solution taking into consideration turbulent numerical solution of
 Kasner universe. Besides although the very early universe can be considered as a perfect conductor with vanishing
 resistivity $\eta$ , we shall consider throughout the paper, a finite resistivity which to begin with shall allow us to
 consider that the dynamo flow is orthogonal to the magnetic sheet where magnetic fields are located. We start from a nonsingular Kasner solution where
 the magnetic field is stretched along the x and y-direction. The solution presents an inflationary
 phase where particles moves along orthogonal direction of x-y plane. The paper is organized as follows: Section II presents
 the cosmic kinematic dynamo solution, where back reaction of the magnetic field on the fluid is not consider. Section III
 presents the conclusions.
 \section{Slow cosmic dynamos in Kasner universe}
 The Kasner solution of Einstein general theory of
 gravitation is given by the metric
 \begin{equation}
ds^{2}=-dt^{2}+t^{2a}dx^{2}+t^{2b}dy^{2}+t^{2c}dz^{2}\label{1}
\end{equation}
Substitution of this metric into the Einstein equations yields the
following constraints on the constants $(a,b,c)$
\begin{equation}
a+b+c=1\label{2}
\end{equation}
\begin{equation}
a^{2}+b^{2}+c^{2}=1\label{3}
\end{equation}
The well-known and so called Kasner nonsingular solution is given by
the constraints $a=b=0$ , $c=1$ which fulfills algebraic expressions
(\ref{1}) and (\ref{2}). With these constraints at hand we may solve
the MHD dynamo equations
\begin{equation}
{\nabla}.\vec{B}=0=\frac{1}{\sqrt{g}}{\partial}_{i}[{\sqrt{g}}g^{ij}B_{j}]=0\label{4}
\end{equation}
where the same equation is fulfilled by the flow velocity $\vec{v}$
and $g_{ij}$ is the Riemannian $3D$ spatial part of the Kasner
metric. The remaining
\begin{equation}
{\partial}_{t}\vec{B}+(\vec{u}.{\nabla})\vec{B}-(\vec{B}.{\nabla})\vec{u}={\eta}{\nabla}^{2}\vec{B}\label{5}
\end{equation}
where ${\epsilon}$ is the resistivity. Since here we are working on
the limit ${\epsilon}=0$ , which is enough to understand the
physical behavior of the fast dynamo, we do not need to worry to
expand the RHS of equation (\ref{5}). The velocity flow is assumed
to be directed to the z-direction while the magnetic field in the
Kasner universe is stretched along the x and y-directions yielding
the following expressions
\begin{equation}
\vec{B}=exp{(pt)}\vec{B_{0}}\label{6}
\end{equation}
where $\vec{B_{0}}=B^{x}\vec{i}+B^{y}\vec{j}$, and
$\vec{u}=u^{z}\vec{k}$ and indices are lowered and raised with the
spatial part of Kasner metric above. Substitution of these
expressions into the MHD equations yields
\begin{equation}
t[{\eta}{\nabla}^{2}\vec{B_{0}}]-[t^{-1}({B_{0}}^{x}{\partial}_{x}+{B_{0}}^{y}{\partial}_{y})c_{1}\vec{k}]=p\vec{B_{0}}\label{7}
\end{equation}
where
\begin{equation}
{u}_{z}=c_{1}(x,y)t\label{8}
\end{equation}
which was obtained from the equation of the incompressible flow
\begin{equation}
{\nabla}.\vec{u}=0\label{9}
\end{equation}
From equation (\ref{7}) one obtain the following set of
equations
\begin{equation}
{\eta}{\nabla}^{2}\vec{B_{0}}=0\label{10}
\end{equation}
which for finite resistivity yields the equation
\begin{equation}
{\nabla}^{2}\vec{B_{0}}=0\label{11}
\end{equation}
where ${\nabla}^{2}$ now is the two-dimensional laplacian and
\begin{equation}
({B_{0}}^{x}{\partial}_{x}+{B_{0}}^{y}{\partial}_{y})c_{1}=0\label{12}
\end{equation}
\begin{equation}
p\vec{B_{0}}=0\label{13}
\end{equation}
this last equation tells us that the cosmic dynamo is marginally
excited,since ${\eta}$ does not vanish with a vanishing p. The
remaining equations yield the following results
\begin{equation}
{\partial}_{z}[tu_{z}]=0\label{14}
\end{equation}
\begin{equation}
\vec{B}={B^{0}}exp[i(x^{2}-y^{2})](\vec{i}+\vec{j})\label{15}
\end{equation}
\begin{equation}
\vec{u}={c_{0}t}e^{i{(x-y)}}\vec{k}\label{16}
\end{equation}
From these expressions we prove the above assertions.
\section{Conclusions}
In the same way one can test for the existence of astrophysical
dynamos in the presence a general relativistic black holes solution
and its accretion discs, here we addressed the problem of the
existence of a cosmic kinematic slow dynamo embbeded into a Kasner
universe which is assumed to be nonsingular from the very beginning.
The future investigation of the nature of this frozen magnetic field
in turbulent anisotropic universes may led us to determine the
magnetic field evolution in the early epochs of the universe taking
into account a more realistic cosmic dynamo model.

\end{document}